\documentclass[journal,transmag]{IEEEtran}
\usepackage{hyperref,color,breqn,multirow}
\usepackage[top=0.7in, left=0.65in]{geometry}
\usepackage{bm}
\usepackage{amssymb}
\usepackage{array}
\usepackage{breqn}
\usepackage{graphicx}
\usepackage{epstopdf}
\usepackage{ulem}
\usepackage{float}
\usepackage{soul}
\usepackage{array,collcell}
\usepackage{booktabs}

\newcommand{\xv}{{\bf x}}

\renewcommand\emph[1]{{\it #1}}

\begin{document}
\title{Variance-based Robust Optimization of {\color{red} a} Permanent Magnet Synchronous Machine}

\author{\IEEEauthorblockN{Piotr A. Putek\IEEEauthorrefmark{1},
E. Jan W. ter Maten\IEEEauthorrefmark{1}, Michael G\"unther\IEEEauthorrefmark{1},
and Jan K. Sykulski\IEEEauthorrefmark{2},~\IEEEmembership{Fellow,~IEEE}}
\IEEEauthorblockA{\IEEEauthorrefmark{1}Chair of Applied Mathematics and Numerical Analysis, Bergische Universit\"at Wuppertal, 42119 Wuppertal, Germany \\ 
}
\IEEEauthorblockA{\IEEEauthorrefmark{2}Electronics and Computer Science, University of Southampton, Southampton SO17 1BJ, U.K.\\ 
} 
\thanks{{\color{black}Manuscript received June 27, 2017; revised June 27, 2017 and June 27, 2017; accepted June 27, 2017. Date of publication xxx xx, 2017; date of current version xxx xx, 2017. 
Corresponding author: P. A. Putek (e-mail: putek@math.uni-wuppertal.de).} 
	
	Color versions of one or more of the figures in this paper are available online at http://ieeexplore.ieee.org.
	
	Digital Object Identifier (inserted by IEEE).}
}
\IEEEtitleabstractindextext{%
\begin{abstract}
This paper focuses on the application of the variance-based global sensitivity analysis for a topology derivative method in order to solve a stochastic nonlinear time-dependent magnetoquasistatic interface problem. To illustrate the approach a permanent magnet synchronous machine has been considered. Our key objective is to provide a robust design of {\color{red}the} rotor poles and of the tooth base in a stator for the reduction of the torque ripple {\color{black} and electromagnetic losses}, while taking material uncertainties into account. Input variations of material parameters are modeled using the polynomial chaos expansion technique, which is incorporated into the stochastic collocation method {\color{black}in order to provide a response surface model}. 
{\color{black}Additionally, we can benefit from the variance-based sensitivity analysis.}
{\color{black}This allows us to reduce the dimensionality of the stochastic optimization problems, described by the random-dependent cost functional.} {\color{black} Finally, to validate our approach, we provide the two-dimensional simulations and analysis, which confirm the usefulness of the proposed method and yield a novel topology of a permanent magnet synchronous machine.}
\end{abstract}

\begin{IEEEkeywords}
Design optimization, Permanent magnet motors, Topology derivative, Robustness, Stochastic processes, Chaos Polynomials, Uncertainty quantification.
\vspace{-4.77mm}
\end{IEEEkeywords}}

\maketitle
\thispagestyle{empty}
\pagestyle{empty}

\IEEEdisplaynontitleabstractindextext
\IEEEpeerreviewmaketitle

\IEEEpubid{\begin{minipage}{\textwidth}\ \vspace{10mm} \\[12pt] \centering 
		0018-9464 \copyright 2017 IEEE. Personal use is permitted, but republication/redistribution requires IEEE permission.\\
		See http://www.ieee.org/publications standards/publications/rights/index.html for more information. (Inserted by IEEE.)
\end{minipage}}

\section{Introduction}
\IEEEPARstart{D}{ue} to the several attractive features, such as high
efficiency and power factor, high torque to weight ratio and
brushless construction, PM synchronous machines have found recently {\color{black}a wide range of applications} in the automotive industry, e.g.,~\cite{IEEEhowto:gieras}. However, in spite of their unquestionable advantages, including also the field weakening capability of 1:5, the Electrically Controlled Permanent Magnet Excited Synchronous Machine
\footnote{\st{The investigation on the development of the ECPSM construction was conducted in the frame of the project
called ''\textit{The Electrically Controlled Permanent Magnet Excited Synchronous Machine (ECPSM) with application
to electro-mobiles}'', supported by the Polish Government under the Grant No. N510 508040.}} 
(ECPSM)~\cite{IEEEhowto:May11}, served here as a case study, 
suffers inherently from the considerable level of the torque pulsation~\cite{IEEEhowto:putek1}. 
{\color{black}This, in turn, may result in the mechanical vibration and acoustic noise and, as a consequence, affect the machine performance.} 
From this perspective, the mitigation of the torque fluctuations is a key issue for the design of a permanent magnet (PM) machine.
 
Yet, in many engineering applications, physical models are very often affected by a relatively large amount of uncertainty~\st{[2]}
, {\color{black} which is caused mainly by the imperfections in manufacturing processes. For example, as a result of the punching and quenching in the assembly of the electric machine, some magnetic properties of the core material can deteriorate~\cite{IEEEhowto:Gmyrek13}. Thus, the solution to the optimization problem may be strongly affected by the uncertainties in both the geometrical and material parameters~\st{[1], [7]}.
In the literature, various methods for suppressing {\color{red}the torque ripple (TR)} have been proposed. In general, they are divided into {\color{black}two main groups of} the deterministic and stochastic optimization methods. Within this context, the topology optimization in particular seems to be a very powerful methodology, e.g.,~\cite{IEEEhowto:Kim2, IEEEhowto:Lim, IEEEhowto:putek1}. Other approaches, which aim to reduce the performance variability, are based on the the statistically originated 
Taguchi method~\cite{IEEEhowto:islam, IEEEhowto:Li04}. {\color{black}Different efficient solutions involve, for example, the use of a gradient indexed method or a perturbation method in order to deal with the 'deterministic' variations 
of some model parameters~\cite{IEEEhowto:Kim10}.} 

From this perspective, there is a need to include uncertainty quantification (UQ) {\color{black}in the modeling phase} to {\color{black} obtain} reliable numerical simulations. For this reason, in {\color{black}our} paper we explore the stochastic collocation method (SCM) combined with the polynomial chaos expansion (PCE)~\cite{IEEEhowto:Xiu07}. {\color{black} This allows for constructing a suitable surrogate - the so-called response surface model - which can be further incorporated into the {\color{black} topology} optimization flow~\cite{IEEEhowto:pute15a,IEEEhowto:pute15b}.}
Our new contribution, in comparison with~\cite{IEEEhowto:pute15b}, is {\color{black}first} to attain the low \st{RT} {\color{red} TR} design of the ECPSM under uncertainties in the transient analysis. {\color{black}Secondly, the methodology for solving a stochastic time-dependent magnetoquasistatic interface problem is based on the variance-based sensitivity {\color{black} analysis}.
The low \st{torque ripple} {\color{red} TR} design with decreased energy consumption profile, found in this way, allows for improving} the PM machine quality by minimizing {\color{black}variations} of the output performance functions.  

\section{Stochastic Forward Problem}
{\color{black}Specifically, for the optimization, when using gradient methods, it is required to consider an efficient computational model. In this respect, the 3D forward problem has been simplified into the 2D FE model, which still yields accurate computational results, \st{[13] and}~\cite{IEEEhowto:pute15b}.} Thus, a 2D model can be described using the magnetic vector potential $A$ for the stochastic quasi-linear system of PDEs, defined on $t \in (0,T]$ with $T>0$ and ${\xv}\,=\,(x,y)^{\top} \in \mathrm{D} \subset \mathbb{R}^2$ as 
\begingroup\makeatletter\def\f@size{9.5}\check@mathfonts
$$
\label{eq:sfp}
	\left\{
  \begin{array}{l l}
    \nabla\cdot\left( \upsilon_{\scalebox{0.55}{{\rm Fe\,}}}\left( \xv,|\nabla A(\theta)|^2, \xi_1\right) \nabla A(\theta)\right) + \sigma(\xi_4)\partial_t A(\theta)= J_i(\xv,t), &\\ 
    \nabla \cdot \left(\upsilon_{\scalebox{0.55}{{\rm air\,}}}\left( \xv, \xi_2\right) \nabla A(\theta)\right) = 0, &\\ 
    \nabla \cdot \left(  \upsilon_{\scalebox{0.55}{{\rm PM\,}}}(\xv, { { \xi_3 }}) \nabla A(\theta) \right) = \nabla \cdot       \upsilon_{\scalebox{0.55}{{\rm PM\,}}}({\xv,  { \xi_3 }}){\rm {\bf M}}(\bf x),\quad \quad \quad \quad ~~\mathrm{(1)} \\ 
  \end{array} \right.
$$
\endgroup 
{\color{black}endowed with} both boundary and initial conditions, where $\theta:=(\xv,t;{\bm \xi}) \in D \times (0,T] \times \Omega$ with the domain $\rm {D}$,
which refers to the sextant region; {\color{black}$\sigma(\xi_4)=\sigma_{\scalebox{0.55}{{\rm Fe\,}}} (1 + \delta_4 \xi_4)$ represents conductivity.}
$J_i({\bf x},t)$, $i=1,2,3$ denotes {\color{red}an excitation current density} and $\Omega$ is a sample space;
${\bf M}$ represents the remanent flux density of the PM, while ${\bm \upsilon}$ denotes the reluctivity.
In particular, a stochastic model for {\color{black} $\upsilon(\cdot;\bm \xi)$ is given by}  
\begingroup\makeatletter\def\f@size{9.5}\check@mathfonts
\setcounter{equation}{1}
\begin{equation} \label{reluctivityrandom}
{\upsilon}(\theta) = \left\{
\begin{array}{ll}
\upsilon_{\scalebox{0.55}{{\rm Fe\,}}}({\bf x},| \nabla A (\theta) |^2) 
(1 + \delta_1 \xi_1) & 
\textrm{for}\;\; {\bf x} \in 
\MakeUppercase{\rm D}_{\scalebox{0.55}{{\rm Fe\,}}} \\
\upsilon_{\scalebox{0.55}{{\rm air\,}}}({\bf x})\,(1 + \delta_2 \xi_2) & 
\textrm{for}\;\; {\bf x} \in 
\MakeUppercase{\rm D}_{\scalebox{0.55}{{\rm air\,}}} \\
\upsilon_{\scalebox{0.55}{{\rm PM\,}}}({\bf x})\,(1 + \delta_3 \xi_3) & \textrm{for}\;\; {\bf x} \in 
\MakeUppercase{\rm D}_{\scalebox{0.55}{{\rm PM\,}}}, \\
\end{array} \right. 
\end{equation}
\endgroup
where ${\bm \xi} =(\xi_1,\xi_2,\xi_3,\xi_4)$
{\color{red}are} assumed to be random variables, defined on some probability space $(\mathrm{\Omega},\mathcal{F},\mathbb{P})${ \color{red}with the event space~$\mathrm{\Omega}$, sigma-algebra~$\mathcal{F}$ and probability measure~$\mathbb{P}$. For the scalings $\delta_j$, see below.}
\vspace{-0.50549mm}
\section{Uncertainty Quantification $\&$ Sobol Decomposition} 

For the uncertainty quantification, 
{\color{black}we consider ${\bf p}({\bm \xi})$ = [$\upsilon_{\scalebox{0.55}{{\rm Fe\,}}}(\xi_1)$, $\upsilon_{\scalebox{0.55}{{\rm air-gap\,}}}(\xi_2)$, $\upsilon_{\scalebox{0.55}{{\rm PM\,}}}(\xi_3)$, $\sigma(\xi_4)$] $\in \Pi$, where }${\xi_j},j = 1,\ldots,4$ are independent and identically uniformly distributed in the interval $[-1,1]^4$ with the constant magnitude {\color{black} $\delta_j = 0.1$,} {\color{black} for $j=1,2,3$ and $\delta_4=0.05$}. Thus, we assume a joint probability density function $g:\Pi \rightarrow \mathbb{R}$, \st{which is associated with $\mathbb{P}$, and that function $u$ is a square integrable function.}
{\color{red}with $\mathbb{P}$, that defines an inner-product $<.,.>_g$ and a space of $L^2_g$ quadratically integrable functions $u$ that can be approximated by} 
\begin{equation}
\label{putek:eq:u}
u \left( {{\bf x},t; {\bf p}} \right)
          \doteq \sum\limits_{i=0}^N  { {{\alpha}}_{i} \left({\bf x}, t \right) } {{ {\Psi}_i}} \left( {{\bf p}} \right)
\end{equation}
with a priori unknown coefficient functions $ {{\alpha}}_i$ and predetermined basis polynomials ${ {\Psi}_i}$
with the orthogonality property $\mathbb{E}\left[{\Psi}_i {\Psi}_j \right]=\delta_{ij}$.
Here, $\mathbb{E}$ is the expected value, associated with $\mathbb{P}$. 
For the calculation of ${\alpha}_i$, 
we applied the SCM with the Stroud-3 formula~\cite{IEEEhowto:pute15b}, which yields 
the solution at each quadrature node ${{\bm \xi }}^{(k)}$, $k=1,\ldots,K$ of the problem ~(1). Next, the multi-dimensional quadrature rule with associated weights $w_{k}$ is used for projecting function $u_k$ into the basis ${ {\Psi}_i}$ \st{as}
\begin{equation}
  \label{last}
  {\alpha}_{i} ({\bf x},t) \doteq \sum\limits_{k=1}^K u \left({\bf x}, {t,\,{\bf p}^{(k)}} 
                                             \right)\Psi_{i} \left( { {\bf p}}^{(k)} \right) {w_{k}} ,
\end{equation}
Finally, the statistical moments are approximated by 
\begingroup\makeatletter\def\f@size{9.5}\check@mathfonts
\begin{equation}
\mathbb{E}_{} \left[ {{ {u}}\left({\bf x},{t;{ {\bf p}}} \right)} \right]\;\doteq\;{ { 
\alpha}}_{0} ({\bf x},t), \mbox{Var}\left[ {{{u}}\left({\bf x}, {t;{ {\bf p}}} 
\right)} \right]\;\doteq\sum\limits_{i=1}^N {\left| {{ { \alpha}}_{i} (t)} 
\right|^{2}},
\end{equation}
\endgroup assuming $\Psi_0 = 1$~\cite{IEEEhowto:Xiu07}. 
{\color{black}Additionally, in order to assess the impact of each uncertain parameter on the output variation, the variance-based sensitivity analysis has been applied. It is based on the Sobol indices, which allow for decomposing the total variance in the form~\cite{IEEEhowto:sobol}
\begin{equation}
\label{soboldecomposition}
S := \sum_{k=1}^{K}S_i + \sum_{1\,\leq i,< j,\leq K}S_{ij} + \ldots + S_{1,2,\ldots,K}.
\end{equation}
Here, $S=\mbox{Var}\left[ {{{u}}\left({\bf x}, {t;{ {\bf p}}}\right)} \right]$ and $S_{i_1 \cdots, i_S}$ denote the total and partial variances, while the PC-based Sobol indices are defined using (\ref{putek:eq:u}) as~\cite{IEEEhowto:sudret2008}
\begin{equation}
\label{sobol_indices}
 SU_{i_1,\dots,i_S }:=S_{i_1, \cdots, i_S}/S,
\end{equation}
where 
\begin{equation}
	S \doteq \sum_{|k|=1}^{K}v_k^2, \quad S_{i_1 \cdots, i_S} \doteq \sum_{|k| \leq K, k \in L} v_k^2
\end{equation}
with $L:= \{k|k_i \geq 1, i \in \{i_1,\dots,i_S \}; k_j=0, j \notin \{i_1,\dots,i_S \}\}$. Moreover, the total sensitivity indices can be easily computed, when mixed terms are involved in the summation, see for example~\cite{IEEEhowto:sudret2008}.
}

\section{Stochastic Optimization Problem} 

{\color{black} For the optimization it is necessary to define some criteria, which allow to asses the design of a PM machine. On the one hand, our objective is to suppress the \st{ripple torque} {\color{red}TR}. Therefore, as the first objective we consider the electromagnetic torque \st{(ET)} {\color{red} $T$}, which in practical applications can be calculated using the virtual work method for a specific position as a partial derivative of the magnetic co-energy {\color{red}$W_{\rm M}$} w.r.t. the angular displacement as $\vartheta$~\cite{IEEEhowto:gieras}
\begin{equation}
\label{et:vm}
T\left(\theta \right):= \frac{\partial W_{\mathrm{M}}}{\partial {\vartheta} }.
\end{equation}
{\color{black} It should be noted that} the electromagnetic torque fluctuations are proportional to the partial derivative of the co-energy stored in a system versus time $\partial_t W_{\rm M}$. On the other hand, our second objective is to minimize electromagnetic losses, defined as 
\begin{equation}
\label{loss:pv}
	P_{\mathrm{avg}}(\theta): = \int_D Q \left(\theta \right) \mathrm{d}{\bf x} + \frac{1}{2} \int_D \partial_t W_M \left(\theta \right)\,\mathrm{d}{\bf x}, 
\end{equation}
where the first term of the above equation refers to the Joule loss, while the second one - in the case of the time periodic excitation - can be interpreted as the energy dissipation due to the irreversible material behaviour (hysteresis). Therefore, 
we formulate the time-dependent stochastic magnetoquasistatic interface problem for the cost functional, which allows for considering both objectives, by incorporating (\ref{et:vm}) into (\ref{loss:pv}) 
\begingroup\makeatletter\def\f@size{9.5}\check@mathfonts
\begin{equation}
\label{loss_P} 
F\left(\boldsymbol{\xi}\right): = \frac{1}{2} \left[\int_{0}^{T}\,\left|{P_{\rm r}(\partial_t A({\bf x},\boldsymbol{\xi})})\right|^2\, +\left|{P_{\rm h}(\partial_t \nabla A({\bf x},\boldsymbol{\xi})})\right|^2\right]\:\mbox{d}t,
\end{equation}
\endgroup
with the first term $P_{\rm r}(\partial_t A({\bf x},\boldsymbol{\xi})): = \int_D \sigma {\partial_t A} \cdot {\partial_t A}\, \mathrm{d}{\bf x}$ and ${P_{\rm h}(\partial_t \nabla A({\bf x},\boldsymbol{\xi})}):=\int_D \partial_t \upsilon {\nabla A} \cdot {\nabla A}\, \mathrm{d}{\bf x}$. 
{\color{black} Here,} we just multiply the hysteresis losses by a factor of two, instead of using the average weighted method. Finally, our stochastic optimization problem with {\color{black}the discreet weak form of eq. (1) as the stochastic constraint} has been defined as 
\begin{equation}
\label{eq:opt-problem}
\begin{array}{lll}
 \min\limits_{\bf p} & : & {\mathbb{E}}\left[ {F({\bf p})} 
\right]
\\ 
  \mathrm{s.t.} & : & {\rm {\bf K}}\left( {\bf p}^{k} \right){\rm {\bf A}}^{k}={\rm {\bf f}}^{k},\;k=1,...,K, \\[0.5ex]  
  && g_1({\bf x})=|{\rm D}_{\rm{{\scalebox{0.55}{FEr}}}}|/|{\rm D}_{\rm{{\scalebox{0.55}{FEr0}}}}|-S_{\rm{{\scalebox{0.55}{FEr}}}} = 0, \\[0.25ex]
  && g_2({\bf x})=|{\rm D}_{\rm{{\scalebox{0.55}{PM}}}}|/|{\rm D}_{\rm{{\scalebox{0.55}{PM0\,}}}}|-S_{\rm{{\scalebox{0.55}{PM\,}}}} = 0, \\[0.25ex]
  && g_3({\bf x})=|{\rm D}_{\rm{{\scalebox{0.55}{FEs}}}}|/|{\rm D}_{\rm{{\scalebox{0.55}{FEs0}}}}|-S_{\rm{{\scalebox{0.55}{FEs}}}} = 0, \\[0.25ex]
 \end{array}
\end{equation}
where $g_1({\bf x})$, $g_2({\bf x})$ and $g_3({\bf x})$ are the deterministic area constraints ${\color{black}|D|}$ related to the initial areas of iron {\color{red}$\left({\rm D}_{\rm FEr0}\right)$} and PM, {\color{red}$\left({\rm D}_{\rm PM0}\right)$} rotor poles and the tooth base in the stator {\color{red}$\left({\rm D}_{\rm FEs0}\right)$}, with some prescribed coefficients, such as $S_{\rm{FEr}}$, $S_{\rm{PM}}$ and $S_{\rm{FEs}}$, respectively. {\color{black}  
The stiffness and mass matrix is denoted by ${\bf K}$.}} 
{\color{black}Furthermore, in order to solve the above-mentioned optimization problem,} {\color{black} we can benefit from the Sobol decomposition and use the partial derivative of the variances (\ref{sobol_indices}) or the partial derivative of the variances with terms of the mixed indices included. In this way, it is also possible to reduce the dimensionality of the optimization problem 
by eliminating \st{these variables, which results in a small range of the variance-based sensitivity} {\color{red} those variables that contribute lowest to the variance-based sensitivity.}
}
\vspace{-0.25cm}
\begin{table}[H]
\begin{center}
  \caption{Main Parameters of the ECPSM Design~\cite{IEEEhowto:putek1}.}
  \label{tab:main-params}       
    \begin{tabular}{lll}
      \hline\noalign{\smallskip}
      Parameter [unit] & Symbol & {\color{red}Value} \\
      \noalign{\smallskip}\hline\noalign{\smallskip}
      Pole number &2p & 12 \\
      Stator outer radius [mm] & $r_{\rm ostat}$ & 67.50 \\
      Stator inner radius [mm] & $r_{\rm istat}$ & 41.25 \\
      One part stator axial length [mm] & $l_{\rm as}$ & 35.0 \\
      Slot openning width [mm] & $w_{\rm oslot}$ & 4.0 \\
      Number of slots & $ns$ & 36 \\
      Number of phases & $m$ & 3 \\
      Permanent Magnet pole & ${\rm NdFeB}$ & 12 \\
      PM thickness [mm] & $t_{\rm m}$ & 3.0\\
      Remanent flux density [T]  & $B_{\rm r}$ & 1.2 \\
      \noalign{\smallskip}\hline
    \end{tabular}
\end{center}
\end{table}
\vspace{-0.75cm}
\section{Numerical Results \& Discussion}
{\color{black}We applied the proposed procedure to design both the rotor poles as well as the base tooth in the stator of the ECPSM machine at on-load state. The main parameters of the PM are given in Table \ref{tab:main-params}. For the simulation of a 2D FEM model {\color{black} at the on-load condition}, \st{the} COMSOL 3.5a has been applied with $I_i({\bf x},t) = 15 [A]$, $i=1,2,3$.
We implemented the variance-based algorithm for the topology optimization using matlab scripts in MATLAB 7.10. The FEM model consists of a triangular mesh with the second order Lagrange polynomials. The areas of the one-pole pair rotor in the initial 2D mode{\color{red}l} were divided into 360 and 480 voxels for the iron and the PM poles, respectively. In addition, the base tooth in the stator consisted of 512 voxels. As a reference model, we used the model depicted in Fig.~\ref{put:top_re}a with the same area of the PM rotor pole as in the optimized structure, shown on Fig.~\ref{put:top}b. Hence, we investigated only the impact of {\color{black}the shape change of} the PM rotor poles, {\color{black}including shape changes of iron poles and tooth base in the stator}, on the performance functions.{~\color{red}To control the geometrical complexity of the various shapes, a version of the level set method – based on the topological derivative as introduced in~\cite{IEEEhowto:Amstutz} – can be used together with the regularization term, as in~\cite{IEEEhowto:Lim},~\cite{IEEEhowto:putek4},~\cite{IEEEhowto:pute15b}.}
To solve a stochastic direct problem, represented by~(1) the SCM with the PC expansion has been applied, where the input variations have been described by a uniform distribution. Due to the assumed probabilistic density function, the Legendre \st{polynomial of order 2 has} {\color{red}polynomials up to order 2 have} been used as the basis. Moreover, the application of the Stroud-3 formula for the ECPSM machine with four random parameters leads to $K = 8$ grid points in the four-dimensional parameter space. The structure of the electric machine, with the optimized shapes of rotor poles and stator teeth, has been found in the 12th iteration of the optimization process{\color{red}, after approximately 8 hours of computing.}
Both structures, before and after the optimization, are shown in Figs.~\ref{put:top} and~\ref{put:top_re}, respectively.}
Subsequently, for reference and optimized topology, the ET is calculated over three cycles. 
The results for the mean and standard deviations of the ET are depicted {\color{black} in} Fig.~\ref{put:top1}. In order to investigate the influence of the robust optimization on the back EMF, we present 
the mean and standard deviations calculated of the back EMF in Fig.~\ref{put:top2}. In addition, the spectral analysis applying FFT to the back EMF, as shown in Fig.~\ref{put:top2}, has been carried out. 
%
\begin{figure}[H]
\begin{center}
\includegraphics[scale=0.27]{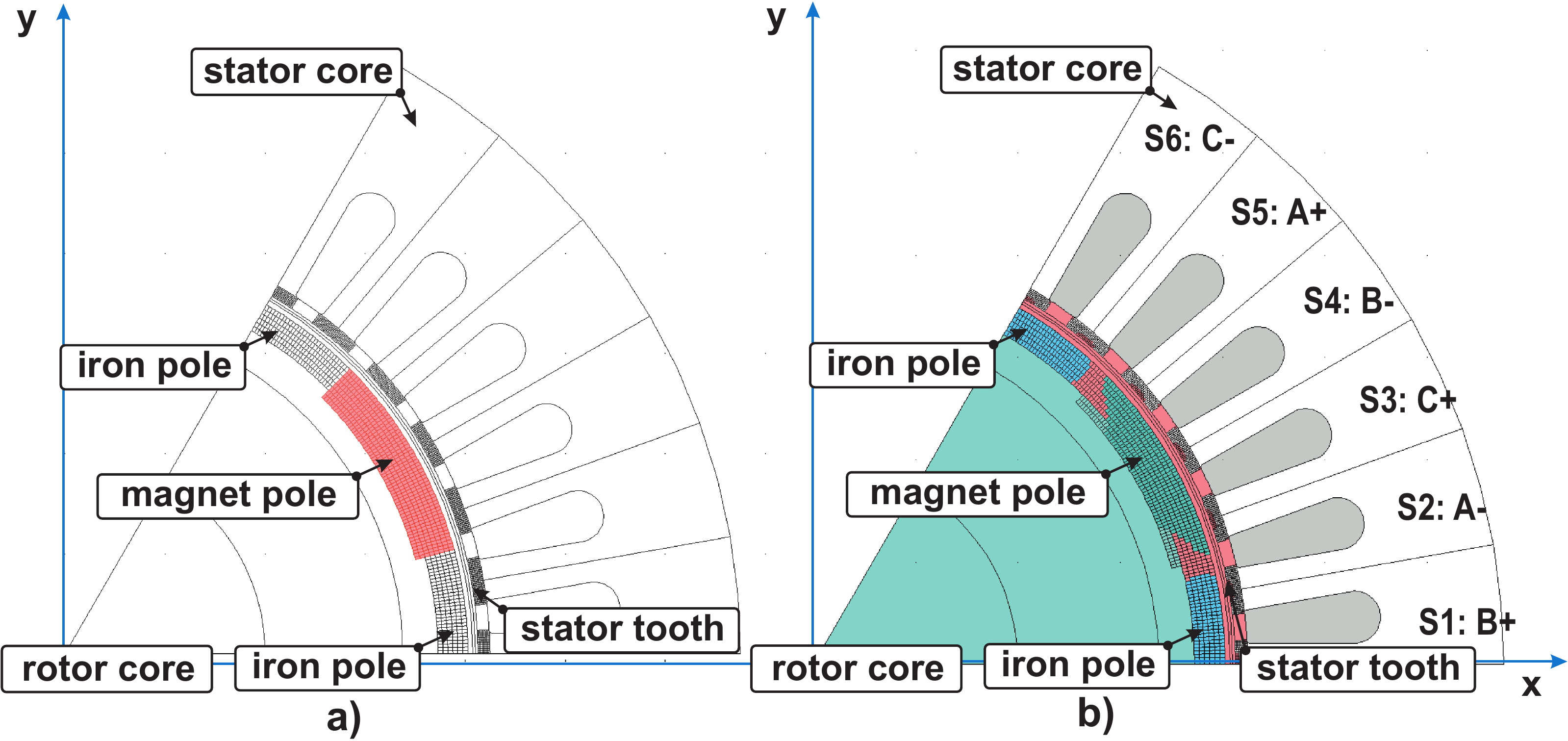}
\end{center}
{
\vspace{-0.5cm}
\caption{\label{put:top}{\color{black} ECPSM topology for (a) initial and (b) optimized configurations.}}
}
\end{figure}
%
\begin{figure}[H]
\begin{center}
\includegraphics[scale=0.23]{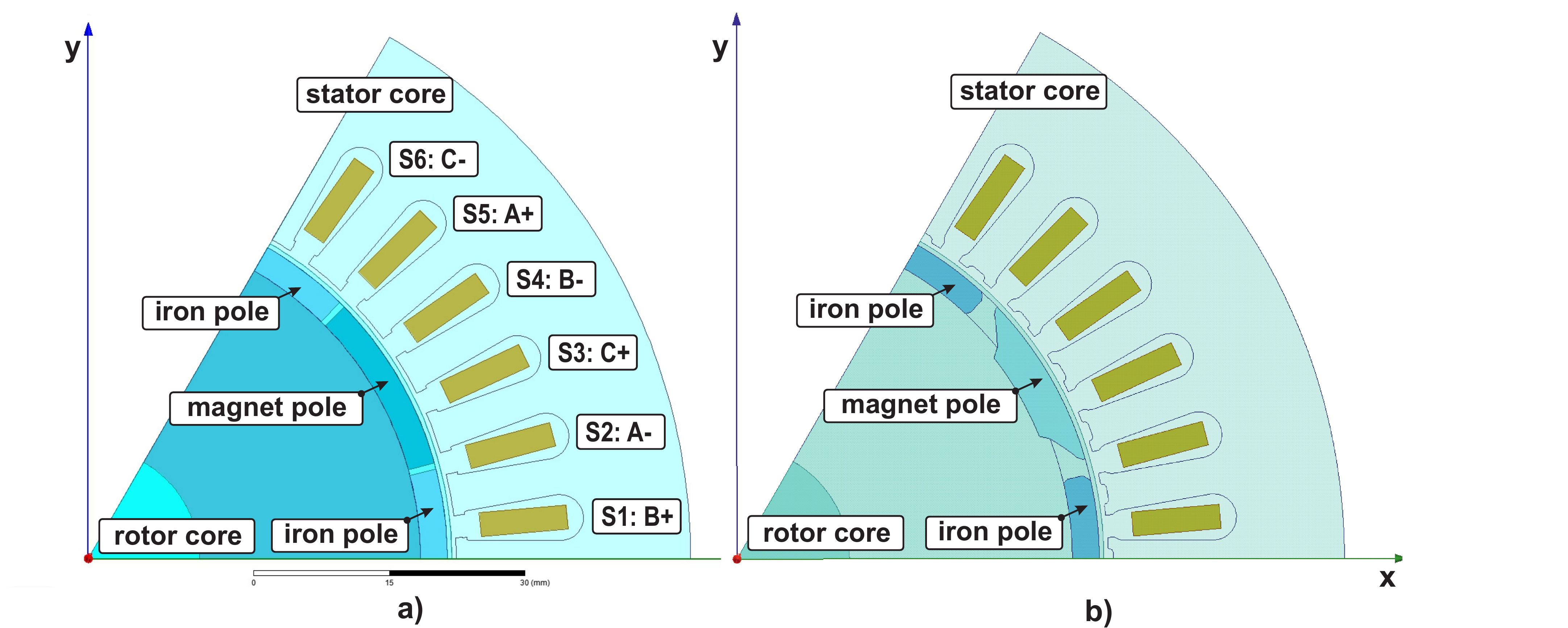}
\end{center}
{
\vspace{-0.5cm}
\caption{\label{put:top_re}{\color{black} ECPSM topology for (a) reference and (b) optimized configurations.}}
}
\end{figure}
%
\begin{figure}[H]
\begin{center}
\includegraphics[width=0.70\linewidth]{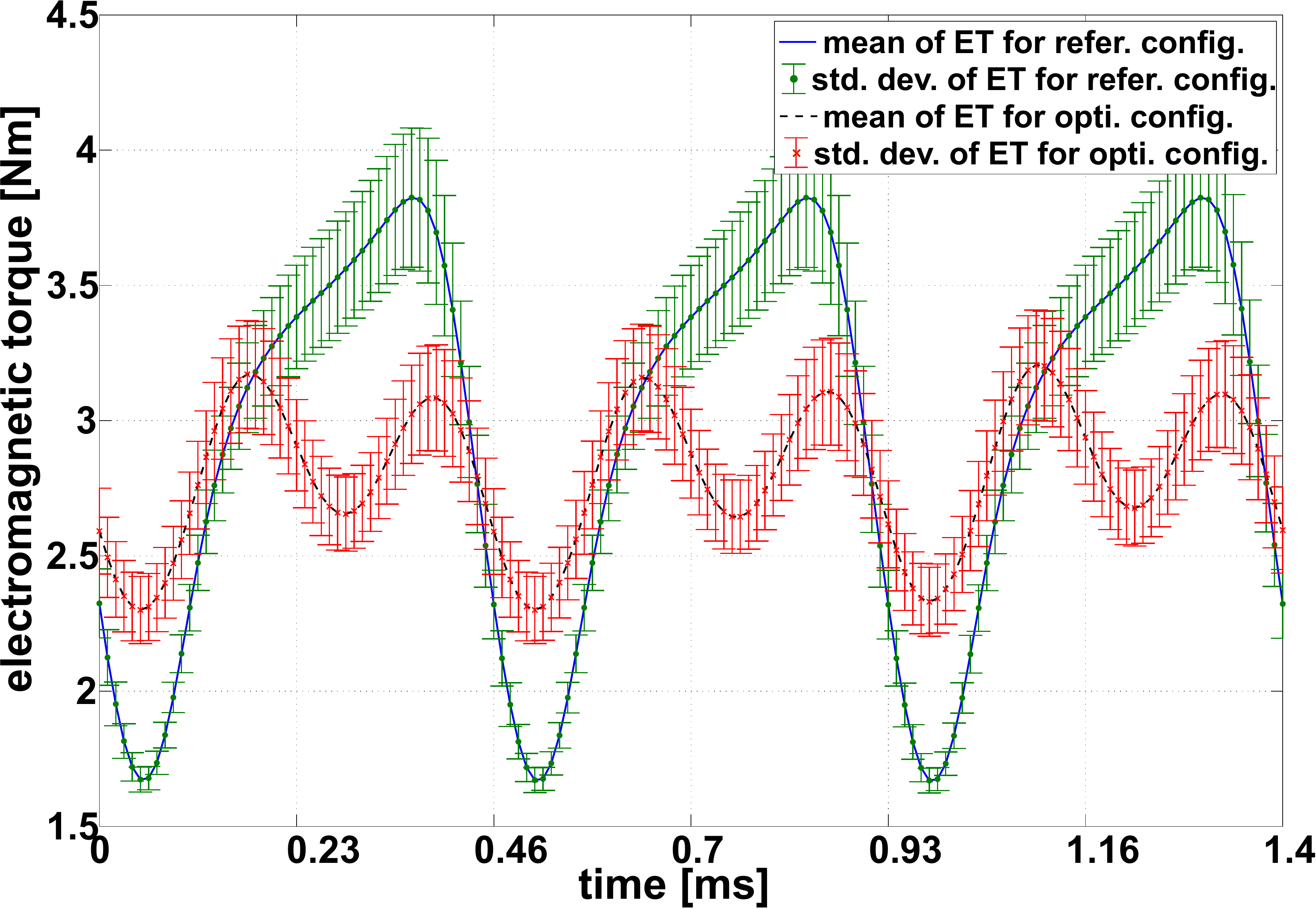}
\end{center}
{
\vspace{-0.5cm}
\caption{\label{put:top1}{{\color{black}Mean and standard deviation \st{of} of the electromagnetic torque (ET) for reference and optimized ECPSM configurations.}}}
}
\end{figure}
%
\begin{figure}[H]
\begin{center}
\includegraphics[width=0.70\linewidth]{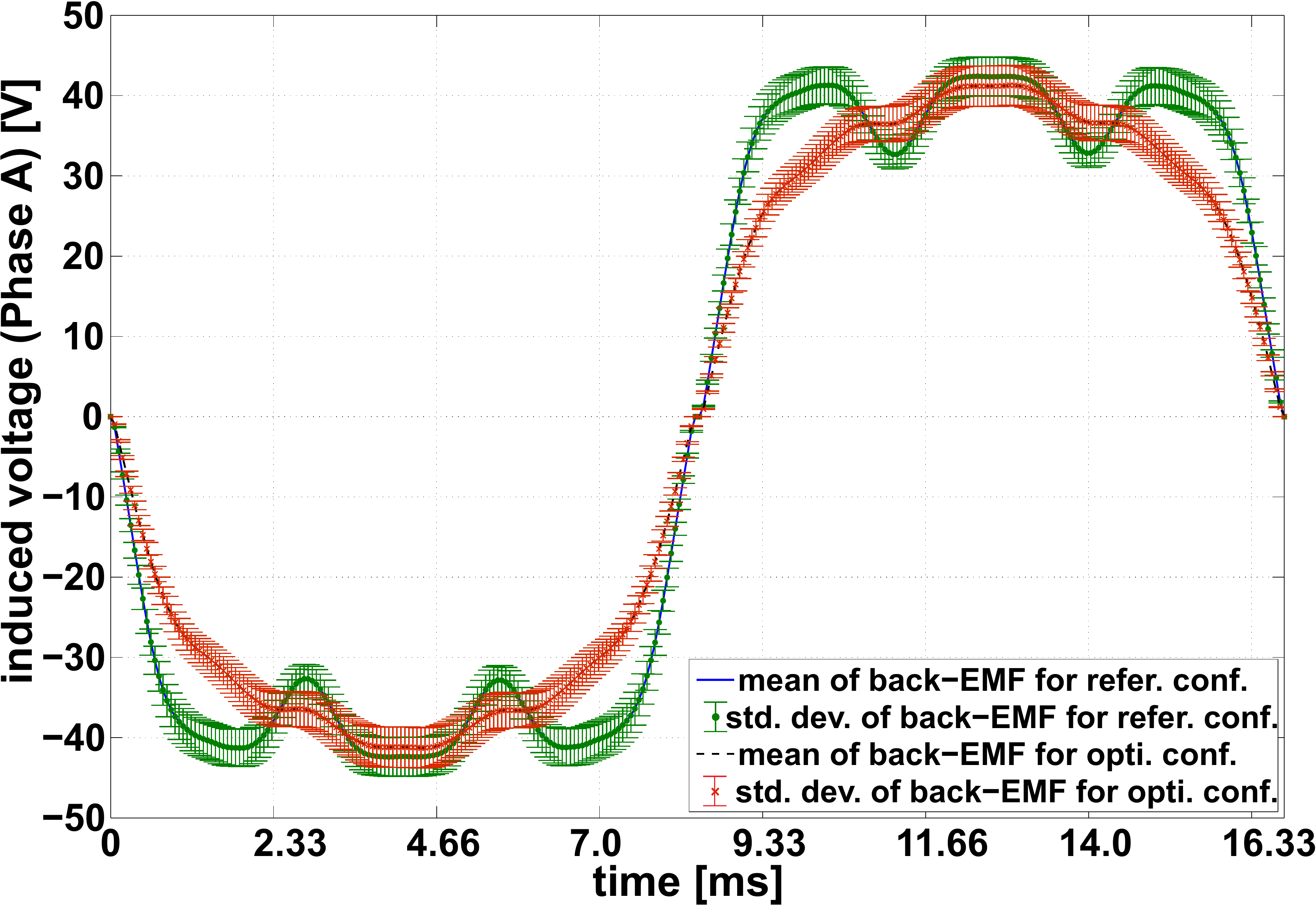}
\end{center}
{
\vspace{-0.5cm}
\caption{\label{put:top2}{{\color{black}Mean and standard deviation of the back EMF for reference and optimized ECPSM configurations.}}}
}
\end{figure}
%
\begin{figure}[H]
\begin{center}
\includegraphics[width=0.70\linewidth]{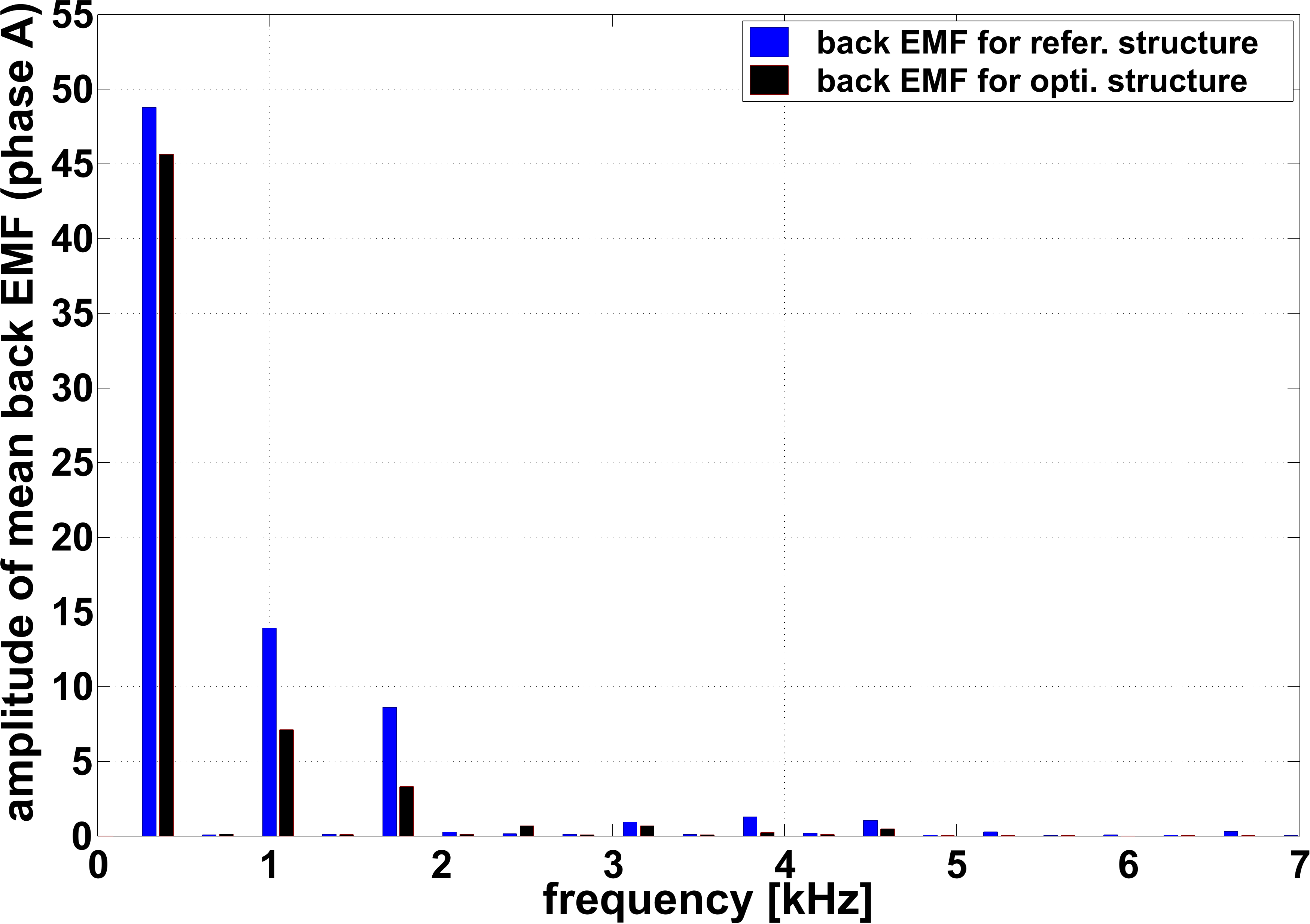}
\end{center}
{
\caption{\label{put:top3}{{\color{black}FFT analysis of the back EMF for reference and optimized ECPSM configurations.}}}
}
\end{figure}
\begin{figure}[H]
\begin{center}
\includegraphics[width=0.70\linewidth]{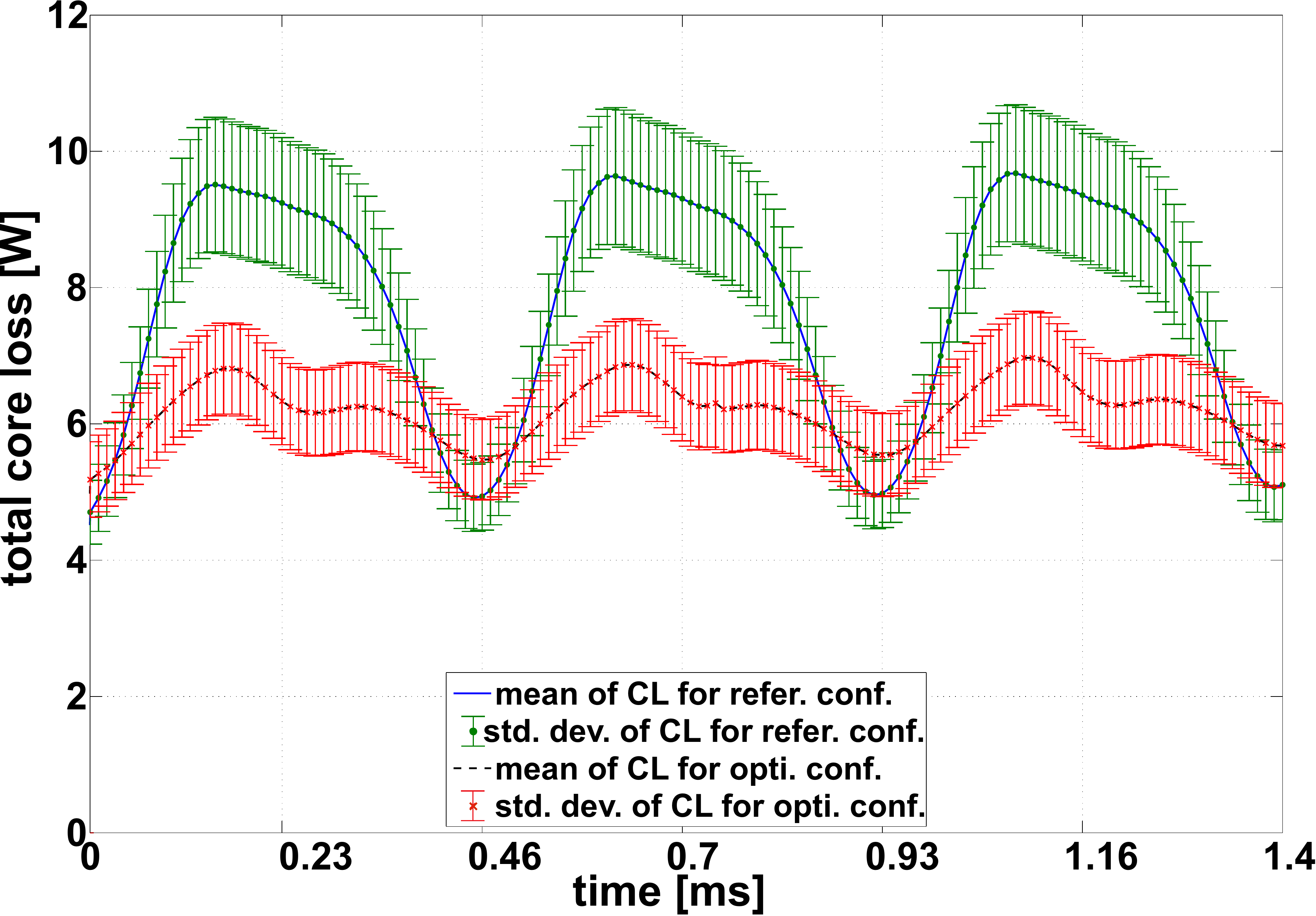}
\end{center}
{
\caption{\label{put:top4}{{\color{black}Mean and standard deviation of of the core loss (CL) for reference and optimized ECPSM configurations.}}}
}
\end{figure}
%
\begin{figure}[H]
\begin{center}
\includegraphics[width=0.70\linewidth]{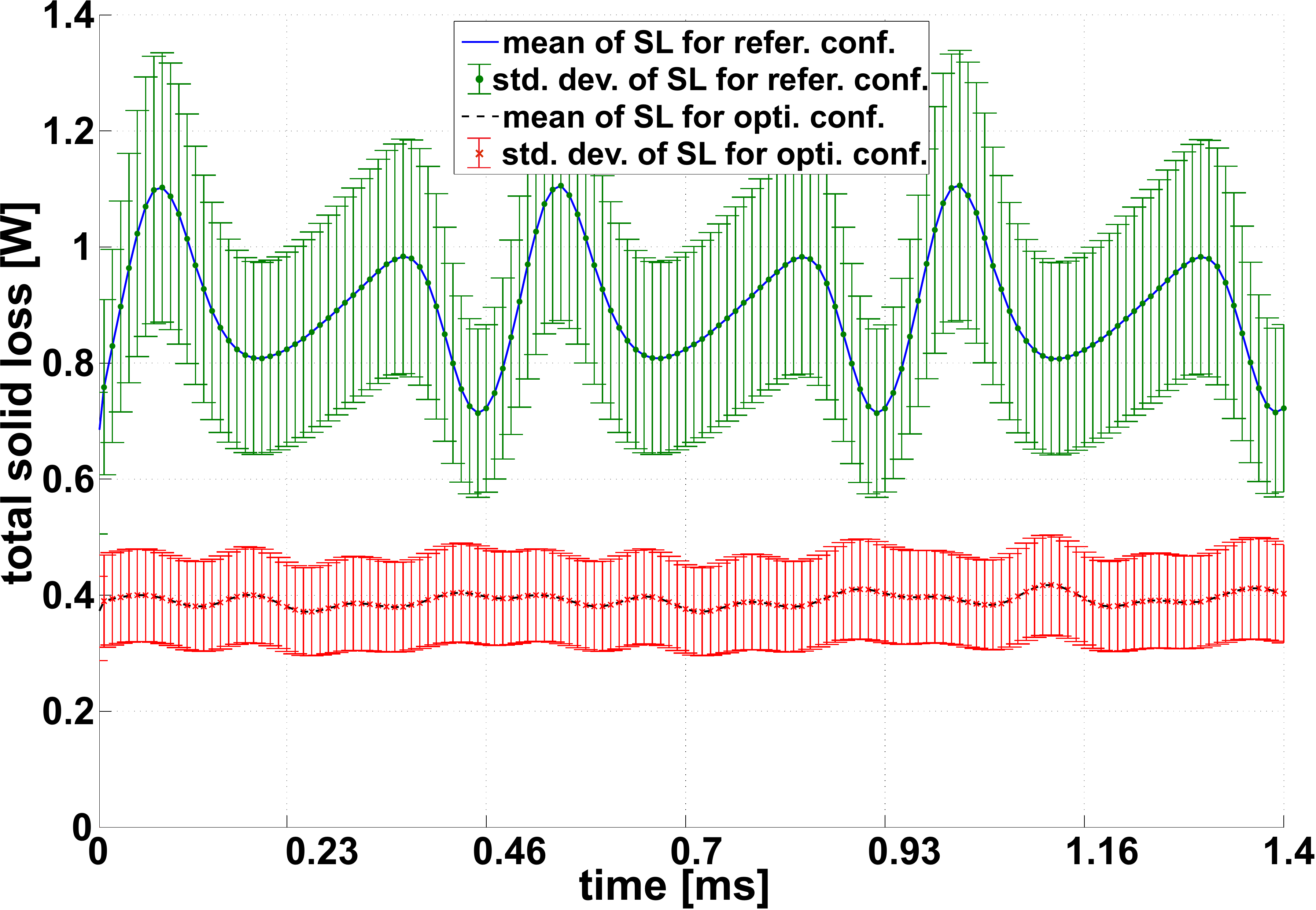}
\end{center}
{
\caption{\label{put:top5}{{\color{black}Mean and standard deviation of the solid {\color{red}(resistive)} loss (SL) for reference and optimized ECPSM configurations.}}}
}
\end{figure}
%
\begin{table}[H]
\caption{Statistics of some Physical Parameters of the ECPSM Reference and Optimized Models at On-load Mode.}
\label{tab:result}  
{\begin{tabular}{llllr@{$\;$}l}
\hline\noalign{\smallskip}
		      & Reference  		& Optimized		& Decrease &\\
      Quantity [unit] & ~topology  & ~topology & ~/increase\hspace*{1ex}&\\
\noalign{\smallskip}\hline\noalign{\smallskip}

      \textbf{Expectation of the ET [Nm]} & & &  \\
      Rectified mean value 	&~2.986 &~2.805 & ~~6.05$\%$ $\downarrow$ \\
      RMS value     			&~3.067 &~2.818 & ~~8.12$\%$ $\downarrow$ \\

      \textbf{Std. dev. of the ET [Nm]} & & & & \\
      Rectified mean value 	&~0.166 &~0.165 & ~~0.71$\%$ $\downarrow$ \\
      RMS value     			&~0.177 &~0.166 & ~~6.25$\%$ $\downarrow$ \\

      \textbf{Expectation of the FL [Wb]} & & & \\
      Rectified mean value 	&~0.014 &~0.013 & ~~5.46$\%$ $\downarrow$ \\
      RMS value     			&~0.016 &~0.015 & ~~6.65$\%$ $\downarrow$ \\

      \textbf{Std. dev. of the FL [mWb]} & & & \\
      Rectified mean value 	&~0.751 &~0.750 & ~~0.15$\%$ $\downarrow$ \\
      RMS value     			&~0.862 &~0.844 & ~~2.13$\%$ $\downarrow$ \\

      \textbf{Expectation of the CL [W]} & & & \\
      Rectified mean value 	&~7.812 &~6.198 & ~20.66$\%$ $\downarrow$ \\
      RMS value     			&~7.984 &~6.212 & ~22.19$\%$ $\downarrow$ \\

      \textbf{Std. dev. of the CL [W]} & & & \\
      Rectified mean value 	&~0.808 &~0.638 & ~21.06$\%$ $\downarrow$ \\
      RMS value     			&~0.827 &~0.638 & ~22.85$\%$ $\downarrow$ \\

      \textbf{Expectation of the SL [W]} & & & \\
      Rectified mean value 	&~0.896 &~0.392 & ~56.27$\%$ $\downarrow$ \\
      RMS value     			&~0.902 &~0.391 & ~56.54$\%$ $\downarrow$ \\

      \textbf{Std. dev. of the SL [W]} & & & \\
      Rectified mean value 	&~0.185 &~0.080 & ~56.51$\%$ $\downarrow$ \\
      RMS value     			&~0.186 &~0.080 & ~56.80$\%$ $\downarrow$ \\

	 \textbf{Mean of others quantities}  & & &\\
      Ripple torque [\%]	&70.31 & 32.13 & ~54.30$\%$ $\downarrow$ \\

\noalign{\smallskip}\hline
\end{tabular}}

\label{symbols}

\end{table}
In practical calculations for the estimation of the electromagnetic losses, the Berttoti model has been applied\cite{IEEEhowto:Bertotti}. The results of this analysis for both configurations have been depicted in Figs.~\ref{put:top3} and~\ref{put:top4}, respectively. Finally, the results for the topology optimization for the quantities such as the electromagnetic torque (ET), the flux linkage (FL), the core loss (CL), and the solid loss (SL) have been summarized in Table \ref{tab:result}. 
\section{Conclusion}
The variance-based topology optimization has been successfully implemented to achieve a low torque ripple as well as reduce electromagnetic losses using a 2D design of an Electrically Controlled Permanent Magnet Excited
Synchronous Machine. \st{Moreover, the same methodology may be extended to a full 3D treatment, however with the inevitable higher computational burden associated with the solution of a 3D stochastic forward problem.} The proposed procedure results in suppressing the mean rms value of the ripple torque (RT) by 54$\%$, and in reducing the mean rms value of the core losses (CL) and and the solid losses (SL) by 22$\%$ and 56$\%$, respectively. 
A drawback of the proposed method is the associated reduction of the rms value of the electromagnetic torque by 8$\%$, and this issue will be addressed in future research.    

\end{document}